\documentclass[aps,prl,twocolumn, superscriptaddress,floatfix]{revtex4-2}
\usepackage{hyperref}
\usepackage{graphicx}
\usepackage[utf8]{inputenc}
\usepackage{amsmath}
\usepackage{mathtools}
\usepackage{algorithm}
\usepackage[noend]{algpseudocode}
\usepackage{verbatim}
\usepackage{amssymb}
\usepackage{color}
\usepackage{mathtools}
\usepackage{bm}
\usepackage{stmaryrd}
\usepackage{subfigure}
\usepackage{tikz}
\usepackage{ifthen}
\usepackage{physics}
\usepackage{ulem}
\hypersetup{
    colorlinks=true,
    linkcolor=blue,
    filecolor=gray,
    urlcolor=blue,
    citecolor=blue,
    }
\usetikzlibrary{calc}
\usepackage{cancel}

\begin{document}

\date{\today}\title{Minimal Pole Representation and Controlled Analytic Continuation of Matsubara Response Functions}
\author{Lei Zhang}
\affiliation{Department of Physics, University of Michigan, Ann Arbor, Michigan 48109, United States of America}
\author{Emanuel Gull}
\affiliation{Department of Physics, University of Michigan, Ann Arbor, Michigan 48109, United States of America}

\begin{abstract}
Analytic continuation is a central step in the simulation of finite-temperature field theories in which numerically obtained Matsubara data is continued to the real frequency axis for physical interpretation.
Numerical analytic continuation is considered to be an ill-posed problem where uncertainties on the Matsubara axis are amplified exponentially.
Here, we present a systematic and controlled procedure that approximates any Matsubara function by a minimal pole representation to within a predefined precision. We then show systematic convergence to the exact spectral function on the real axis as a function of our control parameter for a range of physically relevant setups.
Our methodology is robust to noise and paves the way towards reliable analytic continuation in many-body theory and, by providing access to the analytic structure of the functions, direct theoretical interpretation of physical properties.
\end{abstract}


\maketitle
Quantum field theory simulations at finite temperature are typically performed on the imaginary axis \cite{Mahan13}. In a post-processing step, real-frequency information is obtained via analytic continuation for physical interpretation. Simulations that require continuation range from perturbative calculations \cite{Hedin65,Dahlen05,Phillips14} to lattice \cite{Blankenbecler81} and continuous-time \cite{Gull11} quantum Monte Carlo and lattice QCD  \cite{Asakawa01,Tripolt19,Rothkopf20} simulations, as well as algorithms for the simulation of bosonic systems \cite{Filinov16} including He \cite{Bonisegni96,Vitali10}, supersolids \cite{Saccani12}, and warm dense matter \cite{Bonitz18}.

Due to the ill-conditioned nature of the analytic continuation step \cite{Jarrell96}, a variety of numerical continuation methods have been developed. Among these are Pad\'{e} \cite {Baker96} continued fraction fits  of Matsubara data \cite{Vidberg77,Beach00,Ostlin12,Osolin13,Schott16,Han17}, an interpolation with Nevanlinna functions \cite{fei2021nevanlinna,fei2021analytical}, the Maximum Entropy (MaxEnt) method \cite{Bryan90,Creffield95,Jarrell96,Beach04,Gunnarsson10,Bergeron16,Levy17,Gaenko17,Kraberger17,Rumetshofer19,Sim18}, sparse modeling \cite{Yoshimi19,Otsuki17}, stochastic analytic continuation (SAC) and variants \cite{Shao23,Sandvik98,Mishchenko00,Gunnarsson07,Fuchs10,Goulko17,Otsuki17,Krivenko19}, genetic algorithms and machine learning \cite{Vitali10,Huang22,Yao22}, causal projections \cite{Zhen23} and Prony fits \cite{ying2022pole,ying2022analytic}. In all of these methods, it is difficult in practice to systematically converge the spectral function, even given high-precision Matsubara data.

In this Letter, we revisit the continuation problem from the perspective of a compact low-rank representation of response functions in terms of a pole expansion that approximates Matsubara data within a predetermined precision $\varepsilon$. Remarkably, as we show below, the spectral function systematically converges to the exact answer as the precision of the Matsubara fit is increased. Even `difficult' spectral functions containing both sharp and smooth features at low and at high energies are well approximated. 
\begin{figure}[tbh]
  \includegraphics[width=1.0\columnwidth]{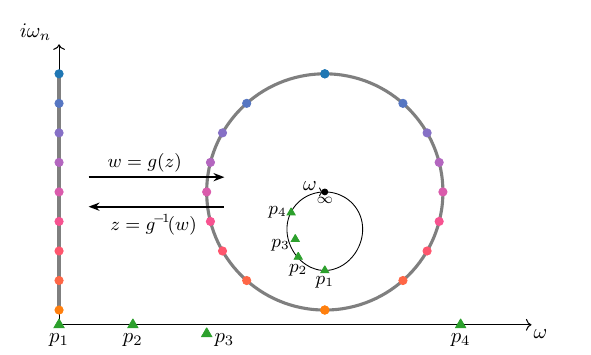}
  \caption{Holomorphic functions $g(z)$  and $g^{-1}(w)$ mapping the complex plane to the unit disk and an interval on the imaginary axis to the unit circle. Also shown are points on and near the real axis as triangles, along with their image under $g$.}
  \vspace{-5mm}
  \label{fig:tri}
\end{figure}

The method is generally applicable to all response functions, including diagonal and off-diagonal fermionic and bosonic response functions of continuous and discrete systems. 
Examining the application of the methodology to data polluted with stochastic noise we find, similarly, that a fit to within the known precision of the input data results in physically reasonable spectral functions that are systematically improved as the uncertainty on the Matsubara axis is reduced. 
\begin{figure}[tbh]
    \centering
\includegraphics[width=1.0 \columnwidth]{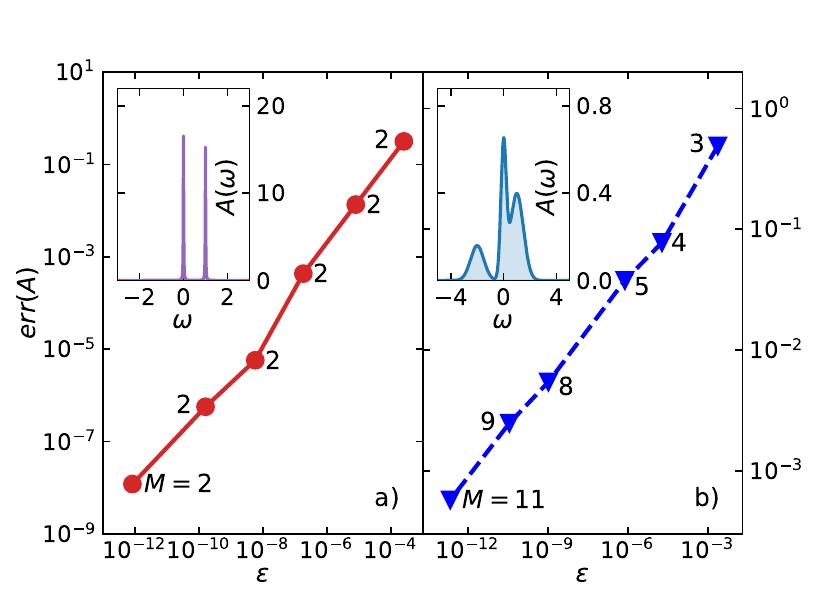}
\caption{Integrated real axis error $err(A)=\int_\mathbb{R}d\omega |A-A_\text{cont}|$ for the discrete (left) and continuous (right) case as a function of control parameter $\varepsilon$. Also indicated is the number of poles $M$. Inset: spectrum $A(\omega)$. Other parameters are $\beta=200$, $n_0=30 \text{ (left)} \text{ and } 0 \text{ (right)}$, $\Delta n = 1$ and $N_\omega = 2001$.}
\label{fig:err_control}
\end{figure}

{\it Theory and Method.}
We construct an approximation of Matsubara data in the upper half of the complex plane by
\begin{align}
\label{eq:pole_representation}
  G(z) = \sum_{l=1}^M \frac{A_l}{z - \xi_l},
\end{align}
where the $\xi_l \in \mathbb{C}$ denote $M$ pole locations in the lower half of the plane and $A_l \in \mathbb{C}$ the corresponding complex weights,  in four steps. First, we approximate Matsubara data on a finite interval of the non-negative imaginary axis using Prony's approximation method \cite{Prony1795,beylkin2005approximation}. Second, we map this interval onto the unit circle using a holomorphic mapping. We then evaluate the moments of the approximated function numerically and use Prony's approximation for a second time to extract a compact representation in terms of pole weights and locations. Finally, we  map the poles back onto the original domain and evaluate the spectral function.

Our input data consists of an odd number $N_\omega = 2N+1$ of Matsubara points $G(i\omega_n)$ that are uniformly spaced, starting from a minimal non-negative frequency $\omega_{n_0}$ with spacing $\Delta n$, i.e., $\{i\omega_{n_0}, i\omega_{n_0 + \Delta n}, \cdots, i\omega_{n_0 + (N_{\omega}-1) \Delta n} \}$.

Prony's interpolation method \cite{Prony1795} interpolates $G_k$ as a sum of exponentials $G_k={\sum_{i=0}^{N-1}} w_i \gamma_i^k$, where  $G_k = G(i\omega_{n_0 + k \Delta n})$, $0 \leq k \leq 2N$, $w_i$ denote complex weights and $\gamma_i$ corresponding nodes.

Prony's interpolation method is unstable \cite{moitra2015super}.
We therefore employ a Prony approximation  \cite{beylkin2005approximation}, rather than an interpolation, of $G$ between $i\omega_{n_0}$ and $i\omega_{n_0 + (N_{\omega}-1) \Delta n}$. For physical Matsubara functions, which decay in magnitude to zero for $i\omega_n \rightarrow i\infty$, only $K\propto \log(1/\varepsilon)$ out of all $N$ nodes in the Prony approximation have weights $|w_i|>\varepsilon$ \cite{beylkin2005approximation}.
More importantly, $K$ significant nodes $w_i$ can be predetermined \cite{beylkin2005approximation} such that the solution to the overdetermined problem for finding weights $w_i$ is stable and yields an accurate solution to the Prony approximation problem
\begin{align}\label{eq:prony_def}
  \left|G_k - \sum_{i=0}^{K-1} w_i \gamma_i^k \right| \leq \varepsilon \text{ for all } 0\leq k \leq 2N 
\end{align}
for a predefined tolerance $\varepsilon>0$ via singular value decomposition. By varying $k$ continuously over the interval $[0, 2N]$, we obtain an approximation of Matsubara data on the continuous interval $[i\omega_{n_0}, i\omega_{n_0 + (N_{\omega}-1) \Delta n}]$. This form of approximation employs a minimum number of exponential sums and is essential for regularizing the problem.

We then apply a holomorphic transform $g(z)$, which is a combination of linear transform and an inverse Joukowsky transform \cite{joukowsky1910konturen} and is illustrated in Fig.~\ref{fig:tri}, to map the complex plane to the closed unit disk $\bar{D}$.
\begin{align}
  \left\{\begin{array}{lll}
    w\!\! &= g(z) &= z_{\rm s} - \sqrt{z_{\rm s}^2 + 1} \text{ with } z_{\rm s} = \frac{z-i\omega_{\rm m}}{\Delta \omega_{\rm h}}  \\
    z\!\! &= g^{-1}(w) \!\!&= \frac{\Delta \omega_{\rm h}}{2}(w - \frac{1}{w}) + i \omega_{\rm m}
  \end{array}
  \right. \; ,
\end{align}
where $\omega_{\rm m} = (\omega_{n_0} + \omega_{n_0 + (N_\omega - 1)\Delta n}) / 2$ is the frequency in the middle of the approximated interval, $\Delta \omega_{\rm h} = (\omega_{n_0 + (N_\omega - 1)\Delta n} - \omega_{n_0}) / 2$ is half of the segment length, 
and the branch of the square root in the first equation is chosen such that $|w| \leq 1$.  The approximated Matsubara interval forms the unit circle, with $g(i\omega_{n_0})=-i$, $g(i\omega_{n_0 + (N_\omega - 1) \Delta n})=+i$, and any other point splits into two copies with identical $y$ values. The real axis is mapped onto a closed contour contained in the unit disk with $\infty$ mapped to the origin.
\begin{figure*}[bth]
\includegraphics[width=1.0\textwidth]{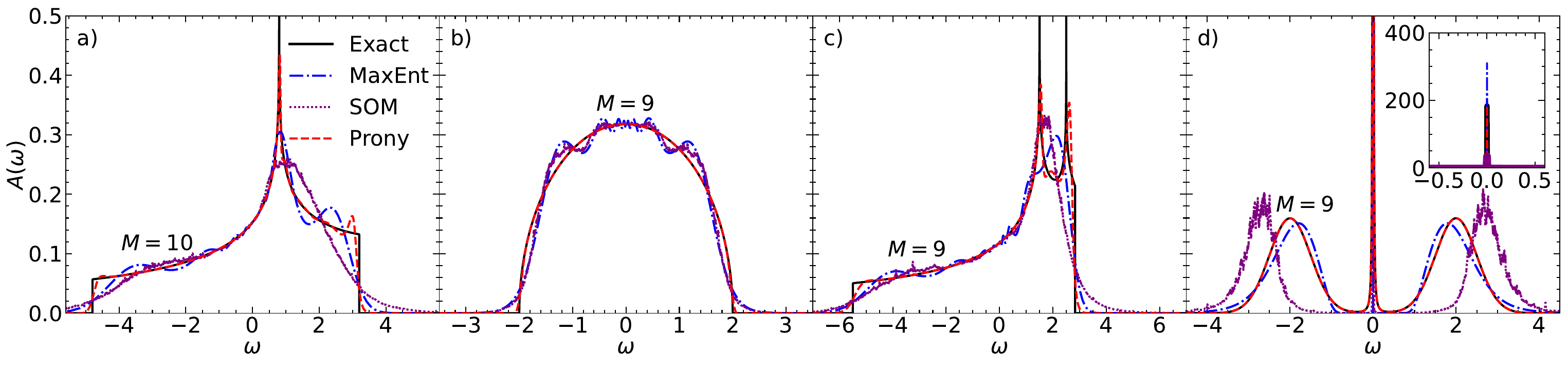}
\caption{Continuation of continuous spectral functions. From left to right: tight binding density of states of 2d square lattice with nearest- and next-nearest neighbor hopping. Semicircular density of states. Tight-binding density of states of the anisotropic triangular lattice. `Kondo'-like spectral function. Shown are the exact input $A(\omega)$ in black, a continuation with Maximum Entropy (blue), SOM (purple), and a Prony fit (this method) in red. Maximum Entropy parameters  fine-tuned to yield best spectra possible.} \label{fig:CLupdate}
\end{figure*}
Since the transformed response function $\tilde{G}(w)$ corresponds to Eq.~\ref{eq:pole_representation}  as $\tilde{G}(w) = G(z)$ and takes the form
\begin{align}
  \tilde{G}(w) = \sum_{l=1}^{M} \frac{\tilde{A}_l}{w - \tilde{\xi}_l} + \text{ analytic part} \; ,
\end{align}
the integrals over the unit circle
\begin{align}\label{eq:contour_int}
  h_k &:= \frac{1}{2\pi i} \int_{\partial \bar{D}} \tilde{G}(w) w^k dw
\end{align}
yield  its moments and, via the residue theorem, pole information \cite{ying2022analytic,ying2022pole}:
\begin{equation}\label{eq:prony_pole}
h_k = \sum_l \tilde{A}_l \tilde{\xi}_l^k,\;k \ge 0 \;.
\end{equation}
Additional simplification of Eq.~\ref{eq:contour_int} yields
\begin{align}\label{eq:hk_int}
  h_k \!=\! 
  \left\{\begin{array}{lll}
    \!\!\!\!\!&\frac{i}{\pi} \int_{-\frac{\pi}{2}}^{\frac{\pi}{2}} G(i (\omega_{\rm m} + \Delta \omega_{\rm h} \sin\theta)) \sin(k+1)\theta d \theta \\
    \!\!\!\!\!&\frac{1}{\pi} \int_{-\frac{\pi}{2}}^{\frac{\pi}{2}} G(i (\omega_{\rm m} + \Delta \omega_{\rm h} \sin\theta)) \cos(k+1)\theta d \theta
  \end{array}\right. \; ,
\end{align} 
for $k$ even and odd, respectively. Using the continuous representation of $G$ obtained in the last step and numerical quadrature, these moments are obtained to high precision. Note that since all $\tilde\xi_l$ lie within the unit circle, the moments $h_k$ decay quickly as a function of $k$ and can be truncated for $h_k \ll \varepsilon$.

Eq.~\ref{eq:prony_pole} forms a second Prony problem. With Eq.~\ref{eq:prony_def}, $M$ significant $\tilde{A_l}$ and $ \tilde{\xi}_l$ are extracted and the resulting poles and weights are recovered as 
\begin{align}\label{eq:location_trans}
  \xi_l &= g^{-1} (\tilde{\xi}_l) = \frac{\Delta \omega_{\rm h}}{2}(\tilde{\xi}_l - \frac{1}{\tilde{\xi}_l}) + i \omega_{\rm m} \;,\\ \label{eq:weight_trans}
  A_l &= {\rm Res}[G(z), \xi_l]  = \frac{\Delta \omega_{\rm h}}{2}(1 + \frac{1}{\tilde{\xi}_l^2}) \tilde{A_l} \;.
\end{align}
Eqs.~\ref{eq:location_trans} and \ref{eq:weight_trans} yield a minimal pole approximation of the form of Eq.~\ref{eq:pole_representation} that is accurate to within $\varepsilon$ and reveals the analytic structure of the function. To evaluate the corresponding spectral function $A(\omega)=-\frac{1}{\pi} \text{Im}G(\omega)$, we evaluate Eq.~\ref{eq:pole_representation} for $\omega$ along $\mathbb{R} + i0^+$. By lowering $\varepsilon$, the  precision can be systematically increased, at the cost of adding additional poles. For the cases examined, this pole representation is much more compact than comparable schemes \cite{Boehnke11,Kananenka16,Gull18,Shinaoka17,Li20,Shinaoka22,Kaye22}
which typically do not yield a systematically improvable representation of the spectral function and may violate the analytic properties of the response function.

Prony's method has previously been used to study the analytic continuation problem~\cite{ying2022analytic,ying2022pole}. The major differences to this work are that Ref.~\cite{ying2022analytic} employs a different approximation procedure, either a causal projection onto a finite real-axis grid or a spline interpolation, and different grids and maps, as well as a different  solution method of the Prony problem. The methodology does not yield the systematic error control observed here.

The supplement \cite{supp} to this paper contains a pedagogical implementation of this procedure that, given a set of Matsubara  points and a tolerance $\varepsilon$, produces a compact representation of the response function and its corresponding spectral function. An open source implementation is also available as part of the Green software package \cite{green,zhang_2024_11520534,PronyACCode}.


{\it Results.} We start our discussion with an examination of the convergence of the spectrum as a function of the error control parameter $\varepsilon$. For a discrete (Fig.~\ref{fig:err_control}a) and continuous (Fig.~\ref{fig:err_control}b) case we define a spectral function $A(\omega)$ on the real axis, transform it to the Matsubara axis, and continue it back to the real axis within precision $\varepsilon$ as $A_\text{cont}$. We then show $err(A)=\int d\omega |A-A_\text{cont}|$ as a function of $\varepsilon$. In striking difference to the `ill-conditioned' nature of a direct analytic continuation, we observe that $A_\text{cont}$ rapidly converges to $A$ as $\varepsilon$ is decreased. The approximation is indeed compact: in the discrete case, only two poles are needed irrespective of the precision. In the continuum case, increasing the precision of the difference of the integral to $10^{-3}$ requires an increase of the number of poles from $M=3$ to $M=11.$
The supplement \cite{supp} contains the precise analytical form of the functions examined along a list of the poles.

We explore the robustness to the number of frequencies and to temperature in the supplement \cite{supp}. We also observe convergent behavior with limited number of frequencies at high temperature, although the analytic continuation problem becomes inherently more difficult. 

In Fig.~\ref{fig:CLupdate}, we analyze the performance of the method for four  continuous noiseless scenarios: A continuous spectral function with sharp band edges and a van Hove singularity, as it is encountered in a 2d tight binding calculation of the square lattice with nearest- and next-nearest-neighbor hopping (left panel); a `semicircular' density of states with square-root singularities as encountered in the non-interacting infinite coordination number Bethe lattice with nearest neighbor hopping (middle panel); a tight-binding band structure of an anisotropic triangular lattice \cite{Yu23}, and a simulated `Kondo' setup with a sharp peak and two side bands (right panel).

We proceed as in Fig.~\ref{fig:err_control} by back-continuing the known function $A(\omega)$ to the Matsubara axis, approximating it with $\varepsilon$ chosen close to machine precision (resulting in $M=10$, $9$, $9$, and $9$), and plotting both $A(\omega)$ and $A_\text{cont}(\omega)$ as a function of frequency $\omega$ together with results from  Maximum Entropy  \cite{Jarrell96,Levy17} and the stochastic optimization method (SOM) \cite{Mishchenko00,Krivenko19}.

All four functions are difficult to analytically continue with standard methods, since they contain both broad and sharp features. The standard methodology of finding the `smoothest' function consistent with input data within some error is not appropriate and introduces artificial `ringing'. While precise knowledge of the location of the band edges and singularities could be used in a Nevanlinna function interpolation \cite{fei2021analytical} followed by a Hardy function optimization \cite{fei2021analytical} to pick the `correct' function out of a Hardy function space, this knowledge is often not available.

The low-rank representation of the Green's function produced by the Prony method provides an unbiased alternative selection criterion that, in this case, is substantially more precise than a smoothness criterion.

\begin{figure}[t]
    \centering
\includegraphics[width=1.0 \columnwidth]{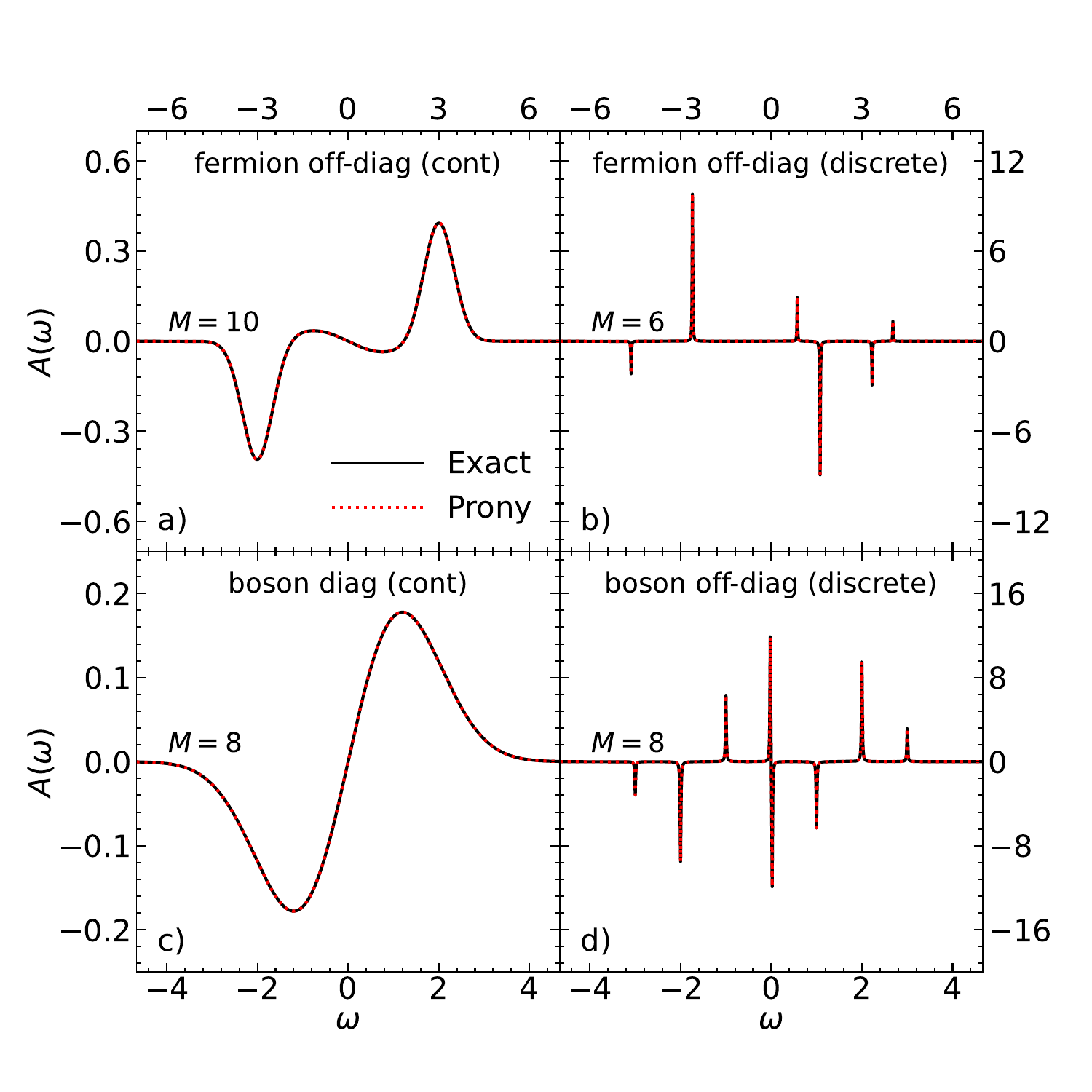}
\caption{Analytic continuation of non-positive spectral functions. Black: Exact input. Red: Continuation. Panel (a/b): off-diagonal continuous (a) and discrete (b) fermion case. (c) diagonal boson case. (d) discrete off-diagonal boson case.}
\label{fig:ver_sys}
\end{figure}
While a fermion Green's function of an operator and its corresponding adjoint corresponds to a positive spectral function \cite{kemper23} whose poles lie in the lower half of the complex plane \cite{fei2021analytical}, response functions of interest also include bosonic, anomalous, and off-diagonal cases which have different analytical properties. Importantly, they may not correspond to a probability distribution, ruling out the straightforward application of Maximum Entropy and related methods. While the issue can be circumvented by continuing related quantities \cite{Gull14,Reymbaut15,Reymbaut17,Nogaki23,Yue23}, the procedure often amplifies errors \cite{Jarrell96}.

The method presented here does not explicitly enforce an analytic structure. It can therefore be applied directly to bosonic, off-diagonal, and anomalous Green's functions as well as to self-energies. As an example we show the off-diagonal part of a continuous fermion spectral function in Fig.~\ref{fig:ver_sys}a; a discrete off-diagonal fermion system in Fig.~\ref{fig:ver_sys}b; a continuous diagonal boson system in Fig.~\ref{fig:ver_sys}c; and a discrete off-diagonal boson system in Fig.~\ref{fig:ver_sys}d. Note that the method for continuous and discrete systems is identical; it is the low-rank representation that places a minimum number of poles very close to the real axis to distinguish sharp (discrete) features from smooth (continuous) ones.

\begin{figure}[t]
    \centering
        \includegraphics[width=1.0 \columnwidth]{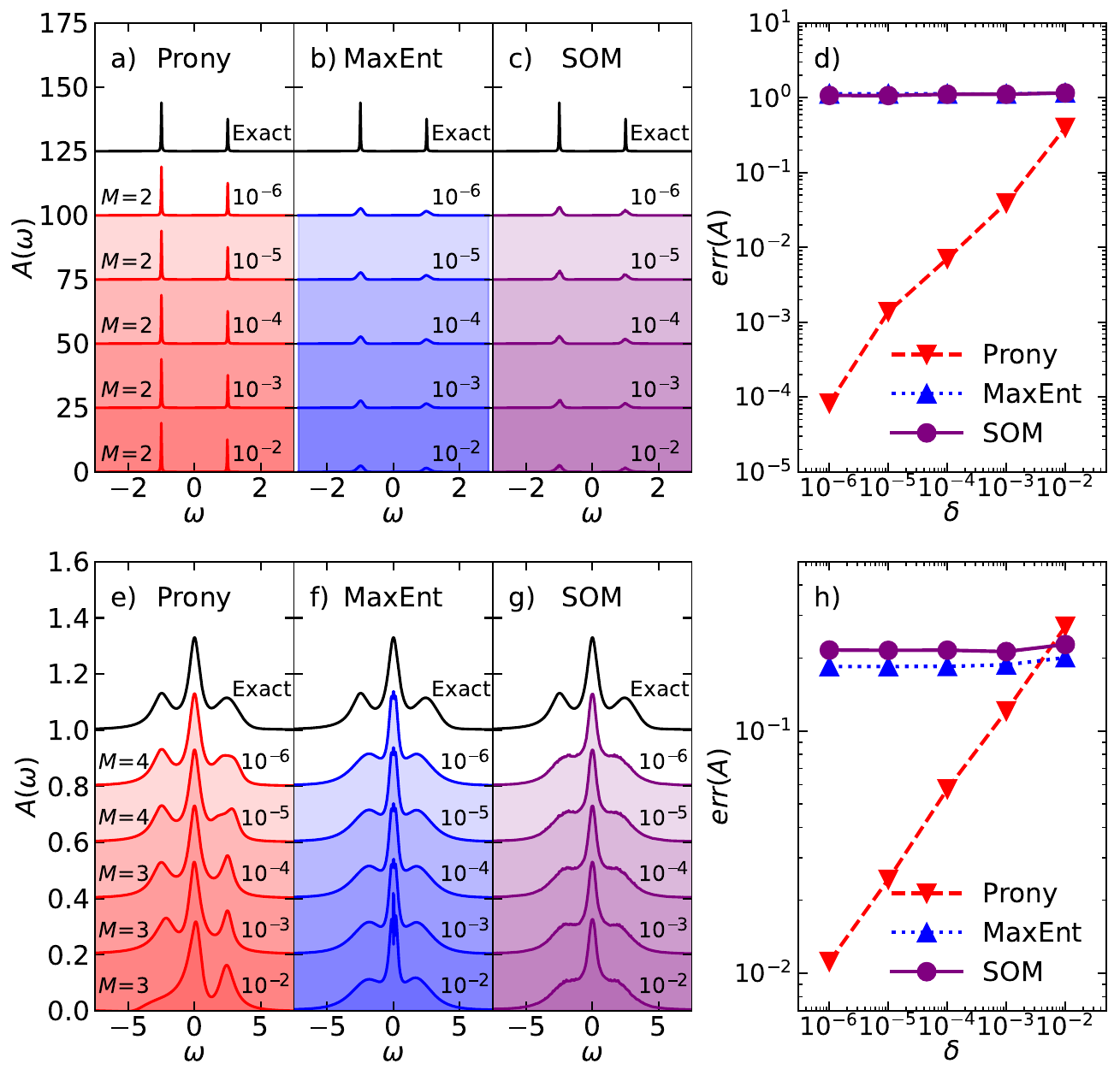}
\caption{Spectral functions for different levels of Gaussian noise $\delta$ on the imaginary axis. Upper panel: discrete case. Lower panel: continuous case. Also indicated is the number of poles $M$.}
\label{fig:noise_resis}
\end{figure}
Analytic continuation is commonly used on noisy Monte Carlo data, where a response function is known only within a given precision. The precision achievable depends very much on the Monte Carlo algorithm and the estimator used but is rarely better than $10^{-5}$, and errors are often (but not always \cite{Wang09}) Gaussian distributed. In that case, we substitute $\varepsilon$ as a proxy for the Monte Carlo error bar.

For a discrete and a continuous scenario, the left panels of Fig.~\ref{fig:noise_resis}
shows the convergence of the spectral function in our method, Maximum Entropy \cite{Jarrell96} and SOM \cite{Mishchenko00} for simulated Gaussian errors with varying magnitude. The right panel shows the integrated error $err(A).$ It is evident that already very loose error tolerance reproduces the main features of the spectrum. As the simulated Monte Carlo errors are decreased, our method rapidly converges to the exact result whereas the spectrum is not recovered in Maximum Entropy and SOM.

In conclusion, we have shown a method to systematically construct low-rank pole approximations to Matsubara response functions of quantum systems and used it to analytically continue spectral functions. We have demonstrated the control of the method in the sense that the error in the real-frequency response functions can be systematically reduced by improving the corresponding Matsubara fit.

We have also demonstrated the wide applicability of the method, including its suitability for diagonal, off-diagonal, fermionic, bosonic, continuous, and discrete response functions and we have examined the convergence in the presence of noise. We note that the same approximation scheme can also be used to model real-frequency response functions a short distance above the real axis, which may be useful in cases where a Matsubara representation is to be avoided entirely.

Apart from analytic continuation, the compact representations introduced here offer a path towards faster numerical and analytical manipulation of response functions, and they offer physical insight by revealing the locations of poles and zeros in the complex plane.

\begin{acknowledgments}
This work was funded by NSF QIS 2310182. L.Z. thanks Yang Yu for the help with the tight-binding model. An open source implementation of the method described here is freely available \cite{supp,PronyACCode,zhang_2024_11520534}
\end{acknowledgments}
\bibliographystyle{apsrev4-2}
\bibliography{reference}
 \end{document}